\documentclass[final,3p,times]{elsarticle}

\usepackage{graphicx}
\usepackage{amssymb}
\usepackage{amsthm}
\usepackage{amsmath}
\usepackage{mathtools}
\usepackage{bm}
\usepackage{lineno}
\usepackage{cleveref}
\usepackage{commath}
\usepackage{parskip}

\newcommand{\ofx}{\left( \bm{x} \right)}

\newcommand{\oft}{\left( t \right)}
\newcommand{\ofxt}{\left( \bm{x}, t \right)}

\newcommand{\ofxyt}{\left( \bm{x},\bm{y}, t \right)}

\renewcommand{\div}{\boldsymbol{\nabla} \cdot}
\newcommand{\grad}{\boldsymbol{\nabla} }
\newcommand{\ddt}[1]{\frac{\partial #1}{\partial t} }
\newcommand{\ave}[1]{\overline{ #1} }
\newcommand{\favre}[1]{\widetilde{#1}}

\newcommand{\mfact}{m}
\newcommand{\Mdot}{\dot{\mathcal{M}}}
\newcommand{\nx}{\grad_{\bm{x}}}
\newcommand{\ny}{\grad_{\bm{y}}}
\newcommand{\eps}{\varepsilon}

\journal{}

\begin{document}

\begin{frontmatter}


\title{Generalised Multi-Rate Models for conjugate transfer in heterogeneous materials}



\author[nott]{Federico Municchi}
\author[nott]{Matteo Icardi}
\address[Nott]{School of Mathematical Sciences, University of Nottingham, University Park,  NG7 2RD, Nottingham, UK}
\begin{abstract}
We propose a novel macroscopic model for conjugate heat and mass transfer between a \emph{mobile region}, where advective transport is significant, and a set of \emph{immobile regions} where diffusive transport is dominant. Applying a spatial averaging operator to the microscopic equations, we obtain a \emph{multi-continuum} model, where an equation for the average concentration in the mobile region is coupled with a set of equations for the average concentrations in the immobile regions. Subsequently, by mean of a spectral decomposition, we derive a set of equations that can be viewed as a generalisation of the multi-rate mass transfer (MRMT) model, originally introduced by Haggerty \& Gorelick \cite{Haggerty1995}. This new formulation does not require any assumption on local equilibrium or geometry. We then show that the MRMT can be obtained as the leading order approximation, when the mobile concentration is in local equilibrium. The new Generalised  Multi-Rate Transfer Model (GMRT) has the advantage of providing a direct method for calculating the model coefficients for immobile regions of arbitrary shapes, through the solution of appropriate micro-scale cell problems. An important finding is that a simple re-scaling or re-parametrisation of the transfer rate coefficient (and thus, the memory function) is not sufficient to account for the flow field in the mobile region and the resulting non-uniformity of the concentration at the interfaces between mobile and immobile regions.  
\end{abstract}


\end{frontmatter}


\section{Introduction}
\label{S:1}
Conjugate transfer in heterogeneous media is of pivotal importance for a wide range of applications ranging from dispersion of contaminants in aquifers \cite{Haggerty1995,Zou2017ModelingSystem,Grisak1981AnDiffusion,Carrera1998,Margolin2003ContinuousSorption} and stagnation/recirculation zones \cite{Zhou2019MassPossible,Crevacore2016,Boutt2006TrappingFracture,Cardenas2007Navier-StokesEddies} to heat transfer in granular media and suspension flows \cite{Forgber2017HeatExchange}, or colloid interface reactions \cite{Ginn2009GeneralizationMemory,Ginn1999}. In all these systems, we are faced with one (or more) flowing fluid exchanging mass or energy with a set of quiescent regions or impermeable inclusions, where diffusion can be assumed to be the dominant transport process. In this work we will refer to the first as \emph{mobile region} and the latter as \emph{immobile regions}. This terminology introduces a classification based on the mathematical modeling of regions rather than their physical meaning, and therefore allows to draw conclusions that are widely applicable to a class of problems. Similarly, we assume that heat and mass transfer processes obey the same governing equations (therefore we do not consider, for example, phase change or other critical phenomena). 

While transport in weakly heterogeneous media can be accurately described using stochastic perturbative approaches \cite{Dentz2003TransportTransfer} (see \cite{Rajaram2002PerturbationAquifers,icardibookchapter,Dagan1989FlowFormations} for an extensive review), typical flow structures and exchange phenomena arising from strong heterogeneities (see for example \cite{dentz2018mechanisms,Webb1996SimulationArchitecture}) can not be captured by low order expansions. In fact, predictions from these methods show significant discrepancies when compared against observations from field experiments \cite{Adams1992FieldAnalysis,Boggs1992FieldDescription}, numerical simulations (for example \cite{Dentz2002TemporalSimulations}) and laboratory experiments \cite{Silliman1987LaboratoryMedia}.

To predict transport in strongly heterogeneous systems, a large number of methods have been developed, the most common of which are:
\begin{itemize}
    \item Integro-differential formulations \cite{Carrera1998} where the mass transfer to the immobile region is represented as the convolution of the concentration with an appropriate memory function over the past history of the system.
    \item The Multi-Rate mass transfer \cite{Haggerty1995}, which consist in modeling the transfer between mobile and immobile regions as a system of first order reactions.
    \item The continuous time random walk \cite{Dentz2003TransportTransfer}, where the movement of solute particles in the heterogeneous medium is represented as random walks in time and space. 
\end{itemize}

Furthermore, it has been demonstrated that these methods are substantially equivalent \cite{Haggerty2000,Dentz2003TransportTransfer,Silva2009} and a unified formulation based on the multi-rate mass transfer has been proposed \cite{Silva2009}. This somewhat arbitrary choice was based on the sound basis that (i) the multi-rate mass transfer is generally more intuitive than the other methods and that (ii) it allows localisation. In the present work, we will add one further reason to motivate such choice: (iii) that the multi-rate mass transfer can be derived from the microscopic equations exactly, and intuitively interfaced with results from homogenisation (see \cite{Pavliotis2008a,Davit2013b} for an extensive review of homogenisation theory).

However, accurate estimation for the closure parameters of the multi-rate mass transfer model is still a largely debated topic. Specifically, the multi-rate mass transfer model of Haggerty $\&$ Gorelick \cite{Haggerty1995} requires a couple of parameters for each first order reaction: 
\begin{itemize}
    \item $\alpha_{\text{HG}}$: the apparent exchange rate coefficient.
    \item $\beta_{\text{HG}}$: the capacity ratio.
\end{itemize}
It was suggested \cite{Haggerty1995,Silva2009} that while these parameters are indeed functions of other variables (like material and geometrical properties) at a fundamental level, they should really be considered as the fundamental coefficients for the model. A formal approach to obtain these coefficient consists in expressing the inter-region transfer as a memory term in the governing equations for the mobile region \cite{Carrera1998,Dentz2003TransportTransfer}. Such term results from the convolution of the accumulation term with a memory function \cite{Ginn2009GeneralizationMemory}, which is then expanded in series of other functions (generally exponentials). The free parameters arising from this operation correspond to the parameters of Haggerty $\&$ Gorelick and they can be evaluated on the basis of analytical solutions for simple geometries \cite{Zhou2019RevisitingFunction}. However, one notorious limitation of such approach is the lack of theoretical basis to describe the dependence of the apparent exchange rate coefficients on the Reynolds number in the mobile region \cite{Gouze2008Non-FickianDiffusion,Zhang2007DoesModelling}. In fact, several studies \cite{Bromly2004Non-FickianSand,Gao2010ADispersion,Pang2002EffectColumns.} showed that an exponential memory function is inadequate to describe the dependence on the flow rate. As a result, more complicated memory functions have been proposed as \emph{ad hoc} solutions \cite{Cvetkovic2012ATransport,Chen1997DescriptionSorption,Russo2010NonidealSediments,Schumer2003FractalTransport}, often based on the breakthrough curves and lacking any sort of physical connection with the underlying geometry or material properties. Therefore, calibration using laboratory experiments or numerical simulations \cite{Li2018,Zhou2019MassPossible} and data fitting are often employed to obtain model parameters in practice.
As a result, current mathematical formulations of multi-rate models still consider (at a macroscopic level) the concentration in the mobile region in equilibrium for what concern the inter-region exchange. 

In this work we propose a novel general derivation of the multi-rate mass transfer model that address the following modeling issues:
\begin{itemize}
    \item[i] Providing a unique way of calculating the model parameters, like a set of equations that can be solved once for a whole class of problems.
    \item[ii] Including the effect of advective transport on the conjugate transfer in a way that is mathematically formal and physically sound.
    \item[iii] Derivation from first principles containing a limited and clear set of assumptions. This with the aim of facilitating any extension in future works.
\end{itemize}

This work is structured as follows. In \cref{S:init} we describe the microscopic equations and the approximations we employ. In \cref{S:ups} we present the upscaling methodology in details and in Section \cref{S:mrmt} we show how the model of Haggerty $\&$ Gorelick can be obtained as a zero-order approximation of our model. In \cref{S:mrmt}, we also present higher order models and we summarise the model parameters in \cref{S:summ}. We conclude in \cref{S:end} with an outlook to future extensions of the current model.

Additional details on the homogenisation procedures can be found in  \ref{S:appendix}.

\section{Assumptions and microscopic equations}
\label{S:init}
\subsection{Heterogeneous domain}
\begin{figure}[h!]
    \centering
    \includegraphics[width=0.7\linewidth]{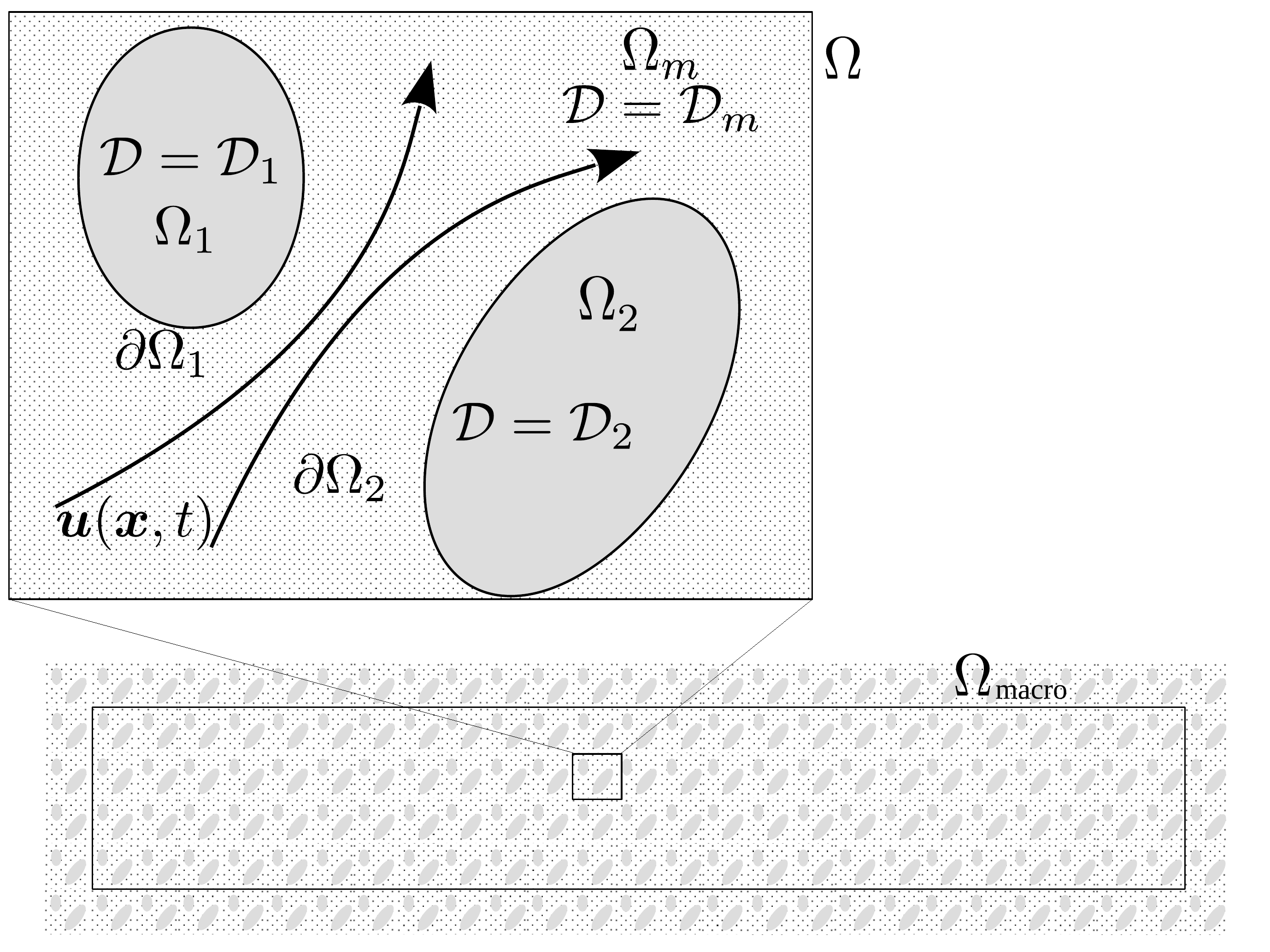}
    \caption{Schematic representation of a domain containing multiple inclusions. A velocity field $\bm{u}(\bm{x},t)$ is defined in the mobile region $\Omega_m$, while only diffusion processes occur in the immobile regions $\Omega_1$ and $\Omega_2$. The diffusion coefficient $\mathcal{D}$ may have different values in each region. This illustration also shows the hierarchy of domains in the multiscale problem. $\Omega_{\text{macro}}$ represents a large collection of similar contiguous REVs, while $\Omega$ is a REV in $\Omega_{\text{macro}}$. Furthermore, $\Omega$ is subdivided into $\Omega_m$ (mobile region), $\Omega_1$, and $\Omega_2$. }
    \label{fig::mrmt_schematics}
\end{figure}

We consider the scenario presented in \cref{fig::mrmt_schematics}. Let us consider a heterogeneous medium composed of a "mobile" region and a number of "immobile" zones. Therefore, $\Omega = \Omega_m \cup_{i=1}^{N_i}\Omega_i$, where $\Omega_m$ is the region occupied by the mobile region and $N_i$ is the number of inclusions.  The mobile region is exchanging mass with the immobile regions through the inclusions' boundaries $\partial\Omega_i$. We also assume that the regions $\Omega_i$ are completely included inside $\Omega$ and that they are disconnected (i.e., they only border on $\Omega_m$).

In the following we will assume that transport within inclusions $\Omega_i$ is dominated by diffusion, while on $\Omega_m$, advection might not be negligible. Thus, we can define a Peclet number:
\begin{equation}
    \label{eq::Peclet}
    \text{Pe} = \frac{ U L}{\mathcal{D}_m}\,,
\end{equation}
where 
$U$ is a characteristic velocity of the fluid, $L$ is a characteristic length and $\mathcal{D}$ is a diffusion coefficient. We will therefore assume that in the immobile regions:
\begin{equation}
    \label{eq::Peclet_cond}
    \text{Pe}_i =  \frac{U_i R_i}{\mathcal{D}_i}\ll 1, \quad  \forall i=1,\dots,N_i 
    \,,
\end{equation}
while no assumption is made on the Peclet number in the mobile region.

\subsection{Microscopic governing equations}
%
We assume that the concentration field $c_m\ofxt$ in the mobile region is obeying the advection-diffusion equation at the microscopic scale:
\begin{equation}
    \label{eq::AD_m}
    \ddt{c_m} + \div \left( \bm{u}c_m - \mathcal{D}_m \grad c_m \right) = 0, \quad \bm{x} \in \Omega_m\,.
\end{equation}
Furthermore, we have $N_i$ diffusion equations for the concentrations in the immobile regions:
\begin{equation}
    \label{eq::D_imm}
    \ddt{c_i} = \mathcal{D}_i \nabla^2 c_i, \quad \bm{x} \in \Omega_i, \quad i=1,\dots,N_i.
\end{equation}
We assume here the immobile diffusion coefficients $\mathcal{D}_i$ to be constant. This can be easily relaxed to smooth or piece-wise smooth coefficients and will be subject of future studies (by decomposing into coupled sub-regions).

At the interfaces $\partial \Omega_i$ we enforce continuity of fields and fluxes:
\begin{equation}
    \label{eq::bc}
    c_i = c_m, \quad \mathcal{D}_i \frac{\partial c_i}{\partial n}  = \mathcal{D}_m \frac{\partial c_m}{\partial n}, \quad \bm{x} \in \partial \Omega_i.    
\end{equation}
This choice of boundary conditions implies that immobile regions do not exchange mass with each other, but they are only connected through the mobile region.

\section{Upscaling methodology}
\label{S:ups}
\subsection{Spatial filtering}

Standard multi-continuum models \cite{Comolli2016} can be obtained from \cref{eq::AD_m} and \ref{eq::D_imm}, by applying a spatial filtering operator to the governing equation for the mobile region using a REV (Representative Elemetary Volume) $\Omega$ as support. 
We will assume that such REVs have a local periodic behaviour or, in other words, that their geometry changes very slowly with $\bm{x}$. Specifically, we assume that the number and geometrical configuration of the immobile regions included in a region $\Omega\ofx$ centred on $\bm{x}$ is essentially equivalent to that of a region $\Omega (\bm{x}+\delta \bm{x})$ for a sufficiently small $\delta\bm{x}$. 
This procedure produces fields that are much smoother than the original ones.

It is important to notice that in our model described by equations \ref{eq::AD_m} and \ref{eq::D_imm}, the diffusive modes are mostly excited by the conjugate transfer with the immobile regions and not by source terms due, for example, to bulk reactions. 

Furthermore, we can consider a macroscopic domain $\Omega_{\text{macro}}$ given by the union of a number of REVs $\Omega$. $\Omega_{\text{macro}}$ is taken sufficiently large to contain a large number of REVs, but sufficiently small to consider all those REVs as equivalent (i.e., disregarding any variation of the REVs geometry and material properties with $\bm{x}$). Therefore, it is possible to interchange between $\Omega$ and $\Omega_{\text{macro}}$ when computing averages without any loss of generality.

\begin{figure}[h!]
    \centering
    \includegraphics[width=0.7\linewidth]{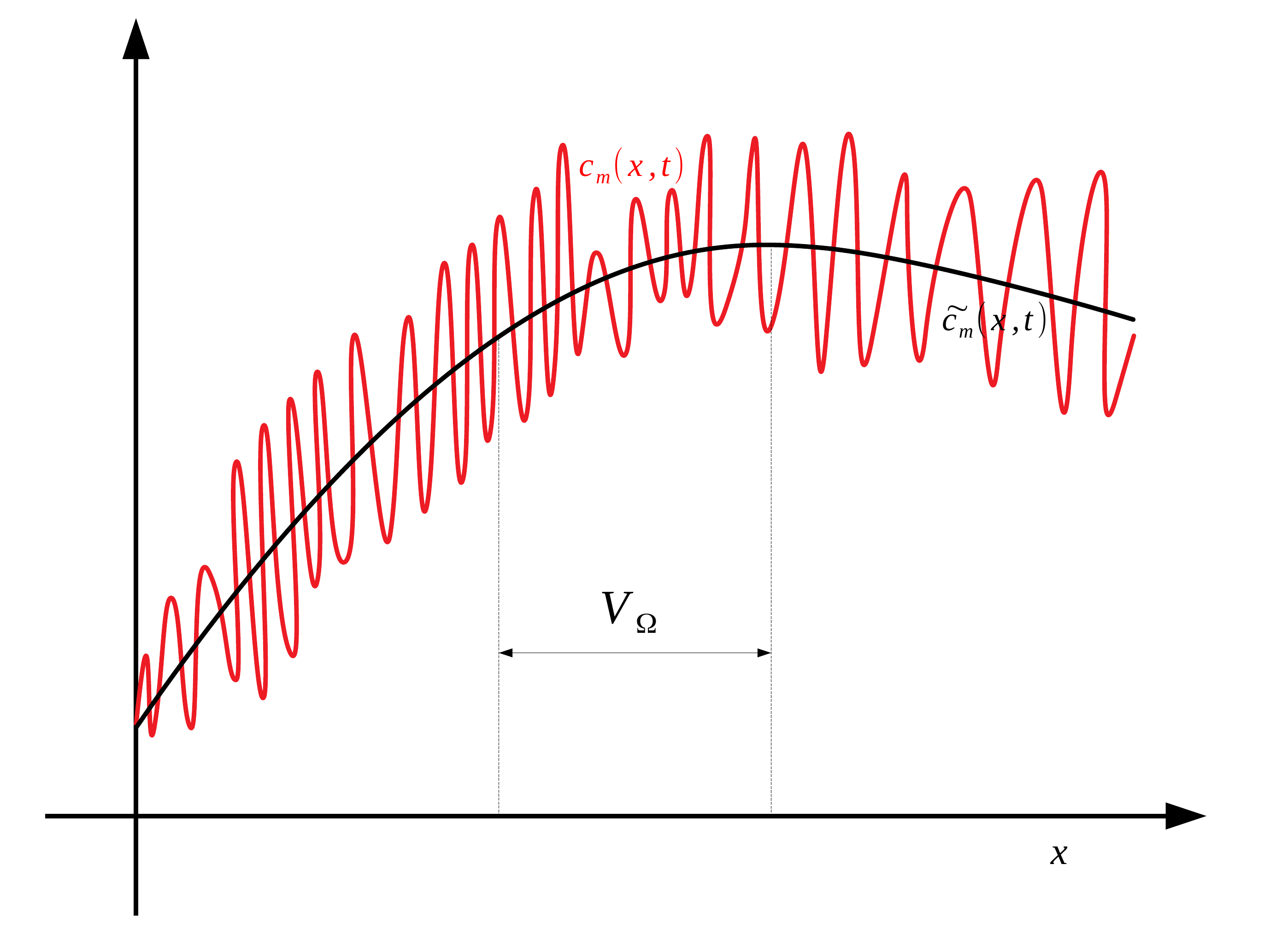}
    \caption{Illustration showing the smoothness properties of $\favre{c_m}$ compared to $c_m$. Here $V$ represents the filter size. $c_m \ofxt$ is filtered at every point $\bm{x}\in \Omega_m$ , so that a value of $\favre{c_m} \ofxt$ is defined at every $\bm{x} \in \Omega$. Notice that $\favre{c_m}$ is defined on the union domain $\Omega$ and not on the perforated domain $\Omega_m$. }
    \label{fig:spatial_filtering}
\end{figure}

Thus, we define the volume average of $c_m$ in the region $\Omega\ofx$ (but it can  be trivially extended to $\Omega_{\text{macro}}$) as the top-hat filter of volume $V = \int_{\Omega}\text{d}V$ centred on $\bm{x}$:
\begin{equation}
    \label{eq::aveVcm}
    \ave{c_m} \ofxt =  \int \limits_{\Omega} K_{\Omega} (\bm{x}-\bm{x}^{\prime} )  c_m (\bm{x}^{\prime} ,t)  \text{d}V = K_{\Omega} * c_m
\end{equation}
Where $K_{\Omega}$ is a filtering Kernel and $*$ is the convolution operator. A typical Kernel that is widely used in fluid dynamics and multiphase flows is the top-hat  \cite{Municchi2016,Municchi2017,Cloete2018}:
\begin{equation}
    \label{eq::tophat_kernelOmega}
     K_{\Omega} (\bm{x}-\bm{x}^{\prime} ) =\frac{1}{V}
     \begin{dcases}
     1 \quad & \forall \bm{x}^{\prime} \in \Omega \ofx \\
     0 \quad & \text{otherwise}
     \end{dcases}
\end{equation}
Where $\Omega \ofx$ is a REV centred on $\bm{x}$. In this formulation both $\ave{c_m}$ and $c_m$ are both functions of the spatial coordinate $\bm{x}$. However, the integral operator results in $\ave{c_m}$ to be much smoother than $c_m$ and therefore we will consider as $\ave{c_m}$ does not depend on space at scales smaller than $\Omega$. Similarly, $V$ is also a slowly varying function of $\bm{x}$. 
We can therefore write an explicit expression for $\ave{c_m}$:
\begin{equation}
    \label{eq::avecm2}
    \ave{c_m} \ofxt = K_{\Omega} * c_m = \frac{1}{V} \int \limits_{\Omega}  c_m (\bm{x}^{\prime},t)\text{d}V
\end{equation}
As commonly done for multiphase and compressible flows, we also define the Favre top-hat Kernel \cite{Municchi2016CPPPO} as the volume average over the mobile region $\Omega_m \ofx$ centred on $\bm{x}$ and of volume $V_m = \int_{\Omega_m}\text{d}V$:
\begin{equation}
    \label{eq::tophat_kernelOmegam}
     K_{\Omega_m }(\bm{x}-\bm{x}^{\prime} ) =\frac{1}{V_m}
     \begin{dcases}
     1 \quad & \forall \bm{x}^{\prime} \in \Omega_m \ofx \\
     0 \quad & \text{otherwise}
     \end{dcases}
\end{equation}

And therefore, the Favre averaged concentration $\favre{c_m}$ can be written as:
\begin{equation}
    \label{eq::favrecm}
    \favre{c_m} \ofxt = K_{\Omega_m} * c_m = \frac{1}{V_m} \int \limits_{\Omega_m}  c_m (\bm{x}^{\prime},t) \text{d}V
\end{equation}

The spatial filtering procedure is illustrated in \cref{fig:spatial_filtering}, where the resulting Favre averaged concentration is much smoother than the original one. 

 Generally, $\Omega$ is itself a function of the coordinate $\bm{x}$ but, since we assume symmetry of $\Omega$ under translation. the integral commutes with spatial derivatives and we will therefore omit its dependence of $\bm{x}$ for the sake of brevity.
such that we obtain the relation:
\begin{equation}
    \label{eq::avefavrecm}
    \ave{c_m} = \beta_m \favre{c_m}\,,
\end{equation}
where we introduced the capacity of the mobile region $\beta_m$ defined as $V_m/V$.
Some authors (for example \cite{Haggerty1995}) define $\beta_m$ as being multiplied by a retardation factor obtained from a re-scaling of the time coordinate. Without loss of generality, we will not consider the retardation factor explicitly.

Thus, the main difference between volume averaging and Favre averaging is that the first is carried over all the space (mobile and immobile regions), while the second is restricted to a particular region. The main reason to introduce such difference, is that it formally leads to equation \ref{eq::avefavrecm}, and thus to the definition of capacity.

\subsection{Multi-continuum formulation}
Assuming that the the immobile regions $\Omega_i$ are fully included into $\Omega$ (i.e., $\partial\Omega \cap  \partial\Omega_i = \emptyset\; i = 1,\dots,N_i$), applying the integral operator \eqref{eq::aveVcm} to \eqref{eq::AD_m} and making use of the Green's theorem we obtain:
\begin{equation}
    \label{eq::mc_1}
    \beta_m \ddt{\favre{c_m}} + \sum \limits_{i=1}^{N_i} \frac{1}{V} \int \limits_{\partial \Omega_i} \mathcal{D}_i \frac{\partial c_i}{\partial n} \text{d}S =
     \beta_m \ddt{\favre{c_m}} + \sum \limits_{i=1}^{N_i}\Mdot_i\oft =
     - \div \mathbf{J}_m\,,
\end{equation}
where we have defined the total average flux in the mobile region:
\begin{equation}
    \label{eq::flux_m}
    \mathbf{J}_m = \ave{\bm{u}c_m } - \ave{\mathcal{D}_m \grad c_m } \,,
\end{equation}
and the average inter-region mass exchange rate for region $i$:
\begin{equation}
	\label{eq::flux_i}
	\Mdot_i\oft=\frac{1}{V} \int \limits_{\partial \Omega_i} \mathcal{D}_i \frac{\partial c_i}{\partial n} \text{d}S\,.
\end{equation}

Since in this work our focus is on the interface exchange, we do not perform an accurate upscaling of $\mathbf{J}_m$, and we will use a simplified expression without any loss of generality:
\begin{equation}
    \label{eq::flux_m_eff}
    \mathbf{J}_{m,\text{eff}} = \bm{u}_{\text{eff}} \favre{c_m} - \mathcal{D}_{m,\text{eff}} \grad \favre{c_m}\,, 
\end{equation}
where $\bm{u}_{\text{eff}} $ and $\mathcal{D}_{m,\text{eff}}$ are the effective velocity and the effective diffusivity in the mobile region. The capacity $\beta_m$ does not appear explicitly into \eqref{eq::flux_m_eff} since it is generally accounted for within the effective parameters.

We then define the Favre averaged concentration in the immobile regions as:
\begin{equation}
    \label{eq::aveci}
    \favre{c_i} \ofxt = \frac{1}{V_i} \int \limits_{\Omega_i}  c_i \ofxt \text{d}V\,,
\end{equation}
where $V_i = \int_{\Omega_i}\text{d}V$ is the volume occupied by region $\Omega_i$. Thus, we integrate \eqref{eq::D_imm} to obtain:
\begin{equation}
    \label{eq::int_imm}
    \ddt{\favre{c_i}}
    =
    \frac{1}{V_i}\int \limits_{\partial \Omega_i} \mathcal{D}_i \frac{\partial c_i}{\partial n} \text{d}S
    =
    \frac{\Mdot\oft}{\beta_i} \,,
\end{equation}
where $\beta_i=V_i/V$ is the capacity of immobile region $i$.
The time derivative in \cref{eq::int_imm} is a partial derivative since $c_i$ depends on $\bm{x}$ at the macro scale (for example, due to the distribution of immobile regions at the macroscale). 
\cref{eq::int_imm}, substituted into \eqref{eq::mc_1}, leads to the multi-continuum equation for the concentration field in the mobile region:
\begin{equation}
    \label{eq::mob_mc}
    \beta_m \ddt{\favre{c_m}} +  \sum \limits_{i=1}^{N_i} \beta_i \ddt{\favre{c_i}}
    = - \div \mathbf{J}_{m,\text{eff}}    
\end{equation}
In \eqref{eq::mob_mc}, we transformed the boundary conditions of the microscopic equation into source terms, one for each immobile region. However, in this formulation $\favre{c_i}$ still needs to be found through an equation that is valid at the microscopic scale and, thus, requires the complete knowledge of the concentration in the immobile region. 

\subsection{Multi-rate mass transfer}
In order to express $\favre{c_i}$ in a closed form that only depends on the geometrical and physical properties of the immobile region (as well as from the boundary value of $c_m$), we perform the following decomposition:
\begin{equation}
    \label{eq::ci_deco}
    c_i\ofxt = \psi_i \ofxt + c_i^{\prime}\ofxt
\end{equation}

Where the function $\psi_i$  satisfies the following equation and boundary conditions:
\begin{equation}
    \label{eq::psi_i}
    \begin{dcases}
     \nabla^2 \psi_i = 0,& \quad \bm{x} \in \Omega_i \\ 
     \psi_i\ofxt = c_m \ofxt,&  \quad  \bm{x} \in \partial\Omega_i
    \end{dcases}
\end{equation}

While $c_i^{\prime}$ is given by:

\begin{equation}
    \label{eq::ciprime}
    \begin{dcases}
    \ddt{c_i^{\prime}} - \mathcal{D}_i \nabla^2 c_i^{\prime}= - \ddt{\psi_i},& \quad \bm{x} \in \Omega_i \\ 
     c_i^{\prime} = 0 ,&  \quad  \bm{x} \in \partial\Omega_i
    \end{dcases}
\end{equation}

Summing \cref{eq::psi_i} and \ref{eq::ciprime}, and using decomposition \ref{eq::ci_deco} gives back \cref{eq::D_imm} with the correct boundary conditions. Due to \cref{eq::psi_i}, $\psi_i$ satisfies the following Gau\ss-Green integral:

\begin{equation}
    \label{eq::Psi_boundInt}
    \int \limits_{\Omega_i} \nabla^2\psi_idV_i = \int \limits_{\Omega_i} \div \grad \psi_i dV_i = \int \limits_{\partial \Omega_i} \frac{\partial \psi_i}{\partial n}\text{d}S_i = 0,
\end{equation}
being $n$ a field normal to $\partial \Omega_i$ and $S_i$ the surface of $\partial \Omega_i$.

Therefore, in our formulation the function $\psi_i$ is simply required to satisfy the non-homogeneous time dependent boundary conditions, while $c_i$ satisfies a non-homogeneous unsteady diffusion equation with homogeneous boundary conditions.

The homogeneous form of \cref{eq::ciprime} leads to an eigenvalue problem following a separation of variables, and can therefore be expressed in series of eigenfunctions without any loss of generality:

\begin{equation}
    \label{eq::ciprime_h}
    c_{i}^{\prime} = \sum \limits_{j=1}^{\infty} c_{ij}^{\prime}\ofxt \phi_{ij} \ofx
\end{equation}

Where $c_{ij}^{\prime}\oft$ are series coefficients that depend on time and on $\bm{x}$ at the macro scale only, while the eigenfunctions $\phi_{ij}\ofx$ carry the dependence on the spatial coordinate at the micro scale and satisfy:
\begin{equation}
    \label{eq::phi_eigen}
    \begin{dcases}
    \mathcal{D}\nabla^2 \phi_{ij} = \lambda_{ij} \phi_{ij},& \quad \bm{x} \in \Omega_i \\ 
    \phi_{ij} = 0 ,&  \quad  \bm{x} \in \partial\Omega_i
    \end{dcases}
\end{equation}
Where $\lambda_{ij}$ is the eigenvalue corresponding to eigenfunction $\phi_{ij}$.

While our decomposition of the spatial dependence in \ref{eq::ciprime_h} may look arbitrary at first sight, in practice it simply mean that there co-exist two problems for the immmobile regions: (i) a local one and (ii) a global one.
The local problem (i) refers to the solution within the single immobile regions and is described by \cref{eq::D_imm} within the REV $\Omega$. The global problem (ii) involves how the fields in the immobile regions vary at a macroscopic scale and how the immobile regions communicate. In our case, the immobile regions are disconnected and therefore, the spatial dependence of $c_{ij}^{\prime}$ is only keeping track of the different initial conditions at the macroscopic scale (since the boundary conditions are homogeneous). Therefore, we will take $c_{ij}^{\prime}$ out of any spatial derivative within the immobile regions, since they are assumed to be negligible.

Both \cref{eq::psi_i} and \ref{eq::phi_eigen} can be made dimensionless by rescaling with respect to a characteristic length of the inclusion $L_i$ and the diffusion coefficient in the immobile region $\mathcal{D}_i$. As a consequence, we can relate the dimensional eigenvalue $\lambda_{ij}$ with a dimensionless eigenvalue:
\begin{equation}
    \label{eq::lambda_dimless}
    \lambda^{\star}_{ij} = \frac{\lambda_{ij}L_i^2}{\mathcal{D}_i}
\end{equation}
Following this rescaling, \cref{eq::psi_i} and \ref{eq::phi_eigen} are not just valid for a particular geometry, but for class of similar geometries. 

Substituting solution \ref{eq::ciprime_h} back into \cref{eq::ciprime} and projecting into $\phi_{ik}$ we obtain:
\begin{equation}
    \label{eq::cijprime_evo}
    \ddt{c_{ik}^{\prime}} = \lambda_{ik}c_{ik}^{\prime} - \frac{1}{A_i}\ddt{} \int \limits_{\Omega_i} \psi_i \phi_{ik}\text{d}V
\end{equation}

Where $A_i = \int_{\Omega_i} \phi_{ik}\phi_{ik}\text{d}V $ is the normalisation factor of the eigenproblem, which depends on the geometry only.

For reasons that will be clear in the next section, we introduce the following definitions:
\begin{equation}
    \label{eq::wij}
    \quad \mfact_{ij} =  \int \limits_{\Omega_i} \phi_{ij} \text{d}V , \quad w_{ij} = \frac{\mfact_{ij}}{A_i}
\end{equation}

\begin{equation}
    \label{eq::cij}
    c_{ij} = \frac{1}{w_{ij}} c_{ij}^{\prime} +   \frac{1}{\mfact_{ij}} \int \limits_{\Omega_i} \psi_i \phi_{ij}\text{d}V
\end{equation}

Substituting into \cref{eq::cijprime_evo} we obtain:
\begin{equation}
    \label{eq::cij_general}
    \ddt{c_{ik}} = \lambda_{ik}\left( c_{ik} - \psi_{ik}\right)
\end{equation}
Where we introduced the projection of $\psi$ into $\phi_{ik}$ scaled over the norm of $\phi_{ik}$:
\begin{equation}
    \label{eq::psi_minus_psistar}
     \quad \psi_{ik} \oft = \frac{1}{\mfact_{ik}}\int \limits_{\Omega_i} \psi_i\ofxt \phi_{ik}\ofx\text{d}V
\end{equation}

Therefore, $c_i$ is now given by:
\begin{equation}
    \label{eq::ci_exp}
    c_i \ofxt = \theta_i \ofxt + \sum \limits_{k=1}^{\infty} w_{ik} c_{ik}\oft\phi\ofx 
\end{equation}

Where:
\begin{equation}
    \label{eq::thetai}
    \theta_i \ofxt = \psi_i\ofxt -  \sum \limits_{k=1}^{\infty} w_{ik}\phi_{ik}\ofx \psi_{ik}\oft
\end{equation}

is the correction function for the immobile region, which accounts for the non-homogeneity of $c_m\ofxt$ at the interface.

\subsubsection{Computation of the exchange rate}

We now compute the Favre averaged concentration in the $i$-immobile region:
\begin{equation}
    \label{eq::ci_favre}
    \favre{c_i}\oft =  \favre{\theta_i} + \sum \limits_{k=1}^{\infty} \beta_{ik} c_{ik}\oft,\quad \beta_{ik} = \frac{w_{ik}m_{ik}}{V_i}
\end{equation}

Where the favre averaged correction function is given by:
\begin{equation}
    \label{eq::thetai_favre}
    \favre{\theta_i} = \favre{\psi_i} \oft- \sum \limits_{k=1}^{\infty} \beta_{ik}\psi_{ik}\oft 
\end{equation}

The terms $\beta_{ik}$ play the role of capacities (or a normalised weighting function) since:
\begin{equation}
    \label{eq::betaiknorm}
     \sum \limits_{k=1}^{\infty} \beta_{ik} = \sum \limits_{k=1}^{\infty} \frac{\left(\int_{\Omega_i} \phi_{ik}\text{d}V\right)^2}{V_i \int_{\Omega_i} \phi_{ik}^2\text{d}V} = 1 
\end{equation}

Equation \ref{eq::betaiknorm} is the so-called \emph{partition of unity} (notice that $\beta_{ik}$ is  generally still functions of the spatial coordinates at the macroscale) and it arises directly from the eigenproblem.  
Recalling \cref{eq::int_imm}, we then obtain an expression for the mobile-immobile exchange rate:
\begin{equation}
    \label{eq::Mdot_general}
    \Mdot_i\oft = \beta_i\ddt{\favre{\theta_i}} + \sum \limits_{k=1}^{\infty}\beta_i\beta_{ik} \ddt{c_{ik}}
\end{equation}

It is important to notice that all the terms involved in the multi-rate transfer can be computed \emph{a priori} by solving a \emph{cell problem}, which consists in solving \cref{eq::psi_i} for $\psi_i$ and the eigenvalue problem \ref{eq::phi_eigen} for each immobile region $i$. However,  $\psi_i\ofxt$ is a non trivial function of $c_m\ofxt$, and in the present formulation its computation requires the solution of equation  \ref{eq::psi_i} for each instant of time. This is clearly not desirable, since it would mean that a numerical algorithm would have to solve \cref{eq::psi_i} at each time step. Furthermore, no information regarding the functional dependence of $c_m$ on the flow rate is provided in the current formulation.
Therefore, we need to introduce some information regarding the micro-structure of $c_m\ofxt$ in order to make any further progress.

\subsection{Representation of $c_m\ofxt$ using homogenisation theory}

So far, our formulation is exact, in the sense that we made no assumption regarding the regularity of the fields and we retained all the terms arising from the volume averaging. However, we still do not have an expression for the concentration field at the interface between mobile and immobile regions since that would require the complete knowledge of $c_m\ofxt$.

In order to give a representation of the spatial variability of $c_m\ofxt$ without having to solve the microscopic unsteady governing equations, one can employ the classical two-scale expansion of homogenisation theory \cite{Davit2013b,bakhvalov2012homogenisation}, and express $c_m$ as:
\begin{equation}
    \label{eq::cm_homo}
    c_m \ofxt = \favre{c_m}\ofxt + \sum \limits_{n=1}^{\infty} \boldsymbol{\chi}_n \ofx: \grad^n\favre{c_m}\ofxt
\end{equation}
where  $\boldsymbol{\chi}_n$ is the \emph{corrector function} corresponding to the $n$-order of the series and $:$ represents the contraction between the corrector and the $n$-order gradient $\grad^n$.

$\favre{c_m}$ and $\grad^n\favre{c_m}$ varies much slower than $\boldsymbol{\chi_n}$ in $\bm{x}$ and can be considered as constant\footnote{They are, in fact, constant at the microscale, when a two-scale hypothesis is introduced.} when plugged into the microscopic equations.
One important feature of corrector tensors is that they provide crucial information on the transport anisotropy. For example, if the flow field is unidirectional, the first order corrector $\boldsymbol{\chi}_1$ will be a vector field  oriented towards the flow direction but (unlike the velocity field) it will not be zero at the interface between mobile and immobile regions. Therefore, employing \cref{eq::cm_homo} allows to reconstruct the interface concentration from the gradients of $\favre{c}_m$ weighted with functions of the transport properties, provided that expansion \ref{eq::cm_homo} is shown to be convergent (which is beyond the scope of this work). We illustrate how the correctors and the above expansion can be obtained using homogenisation theory in \ref{S:appendix}.

As a side note, we mention that homogenisation can be also employed to obtain an expression for $\mathbf{J}_{m,\text{eff}}$ \cite{Allaire1992a,Auriault1995} and it is therefore synergic to the current problem. Furthermore, homogenisation theory can also be employed in place of volume averaging to derive dual porosity models \cite{Arbogast1990}.

To introduce the information provided by the corrector equation into our problem, we can expand $\psi_i$ in a similar fashion:
\begin{equation}
    \label{eq::psii_exp}
    \psi_i \ofxt = \favre{c_m}\ofxt + \sum \limits_{n=1}^{\infty} \boldsymbol{\Psi}_{in} \ofx : \grad^n\favre{c_m}\ofxt
\end{equation}
where the functions $\boldsymbol{\Psi_n}$ are coupled with the correctors $\boldsymbol{\chi}_n$ at the interface and satisfy (owing the linearity of equation  \ref{eq::psi_i}) :
\begin{equation}
    \label{eq::Psi_in}
        \begin{dcases}
         \nabla^2 \boldsymbol{\Psi}_{in}\ofx = 0,& \quad \bm{x} \in \Omega_i \\ 
         \boldsymbol{\Psi}_{in} \ofx = \boldsymbol{\chi}_{n} \ofx,&  \quad  \bm{x} \in \partial\Omega_i
        \end{dcases}
\end{equation}

These are a set of partial differential equations for tensors of rank $n$. Notice that $\boldsymbol{\Psi}_{in}$ satisfies a  boundary integral relation similar to \cref{eq::Psi_boundInt}. We can now substitute expansion \ref{eq::psii_exp} into $\favre{\theta_i}$ to obtain:
\begin{equation}
    \label{eq::favreci_homo}
    \favre{\theta_i}= \sum \limits_{n=1}^{\infty} \left(\favre{\boldsymbol{\Psi}_{in}} - \sum_{k=1}^{\infty}\beta_{ik} \int \limits_{\Omega_i} \boldsymbol{\Psi}_{in}\phi_{ik}\text{d}V \right):\grad^n\favre{c_m} = \sum \limits_{n=1}^{\infty}\left(\favre{\boldsymbol{\Psi}_{in}} - \sum_{k=1}^{\infty}\beta_{ik} \boldsymbol{\Psi}_{ink} \right):\grad^n\favre{c_m} = \sum \limits_{n=1}^{\infty} \boldsymbol{\Theta}^{\star}_{in}:\grad^n\favre{c_m}
\end{equation}
where we introduced $\boldsymbol{\Theta}^{\star}_{in}$ as the internal corrector tensor of rank $n$ for immobile region $i$, which accounts for the internal effects of the spatial variability of $c_m\ofxt$ at the interface.
This formulation shows that, when we assume $c_m\ofxt=\favre{c_m}$ at the interface, then $\favre{\theta_i} = 0 $, which means that no correction is necessary.

Expansion \ref{eq::psii_exp} can now be substituted into the evolution equation for $c_{ik}$, leading to:
\begin{equation}
    \label{eq::ddtcik_exp}
    \ddt{c_{ik}} = \lambda_{ik}\left(c_{ik} - \favre{c_m} - \sum \limits_{n=1}^{\infty}\boldsymbol{\Psi}_{ikn}:\grad^n\favre{c_m}\right)
\end{equation}

\section{Governing equations of the generalised multi-rate transfer model}
\label{S:mrmt}
We can now write down a set of equations for the Generalised Multi-Rate Transfer (GMRT) model which can be closed using a set of parameters corresponding to different geometries. When a specific geometry is selected, such parameters are constants or are simple function of geometrical and material properties through a rescaling (as for the eigenvalue $\lambda_{ik}$ =$\lambda_{ik}^{\star}\mathcal{D}_i/L_i^2$). The full system of equations is:
\begin{equation}
    \label{eq::mrmt_general}
    \mbox{(GMRT-TS)}\qquad 
    \begin{dcases}
    \beta_m \ddt{\favre{c_m}} +  \sum \limits_{i=1}^{N_i} \beta_i\ddt{}\left(\sum \limits_{n=1}^{\infty} \boldsymbol{\Theta}^{\star}_{in}:\grad^n\favre{c_m} + \sum \limits_{k=1}^{\infty} \beta_{ik} c_{ik}\right)= - \div \mathbf{J}_{m,\text{eff}},&   \\
        \ddt{c_{ik}} = \lambda_{ik}\left(c_{ik} - \favre{c_m} - \sum \limits_{n=1}^{\infty}\boldsymbol{\Psi}_{ikn}:\grad^n\favre{c_m}\right),& \quad \begin{matrix} i=1,\dots,N_i,\\k=1,\dots,\infty,\\\end{matrix}
    \end{dcases}
\end{equation}

In this system  (GMRT-TS) mixed time-space derivatives are present. In the next section, these will be replaced to obtain a more convenient form. Nevertheless, GMRT-TS is exact as long as $c_m\ofxt$ can be expanded using the corrector \cref{eq::cm_homo}, and the series is convergent. In practical applications, one would also truncate both the series in $n$ and $k$ to achieve the desired accuracy or retain only a certain number of terms. In that case, some considerations on the approach of the series to convergence are required. However, if the macroscopic field $\favre{c}_m$ is sufficiently regular it is possible to obtain a good approximation just with the first order corrector $\boldsymbol{\chi}_{1}$ \cite{Davit2013b}.

A key feature of the current formulation is that accounts for the non-uniform distribution of the concentration field at the interface from a microscopic perspective and shows how this can be upscaled to a macroscopic set of equations. Surprisingly, this does not lead to a new exchange rate (which is equivalent to the eigenvalue of the homogeneous problem $\lambda_{ik}$), but instead to an additional term in the equations for $c_{ik}$ and a new rate term. These terms lead to mixed and potentially high order derivatives in the governing equation for the mobile concentration. However, in practical applications one rarely goes beyond a second order corrector and therefore this does not alter the order of the differential equation.

\subsection{A note on the truncation of the multi-rate series: equilibrium modes}

Clearly, practical applications require the  multi-rate series to be truncated at some value $k_{\text{max}}$ corresponding to $\beta_{ik_{\text{max}}}$, $\lambda_{ik_{\text{max}}}$, $c_{ik_{\text{max}}}$, and $\boldsymbol{\Psi}_{ik_{\text{max}}n}$. While previous works enforced $\sum_{k}\beta_{ik}=1$ by rescaling the capacities \cite{Silva2009}, here we propose a more accurate and rigorous approach based the above mathematical derivation of the GMRT.

It is easy to see that the truncated modes $k>k_{\text{max}}$ will approach equilibrium faster, since they correspond to small perturbations inside the immobile region and to larger eigenvalues. They can be therefore assumed to be in equilibrium if $k_{\text{max}}$ is chosen sufficiently large, according to the physics of the problem.  Therefore, one can write:
\begin{equation}
    \label{eq::im_highk}
    \ddt{c_{ik}} = 0 \quad \mbox{for}\, {k>k_{\text{max}}}
    \Longrightarrow c_{ik} = \favre{c_m} - \sum \limits_{n=1}^{\infty}\boldsymbol{\Psi}_{ikn}:\grad^n\favre{c_m}
    \quad
    \mbox{for}\, {k>k_{\text{max}}}
\end{equation}

Furthermore, the corresponding eigenfunctions $\phi_{ik}$ will be highly oscillating for $k>k_{\text{max}}$ in contrast with the corrector $\boldsymbol{\Psi}_{in}$, which we choose to be a smooth function (recall equation \ref{eq::psi_i}). Therefore the projection of $\boldsymbol{\Psi}_{in}$ over $\phi_{ik}$ will be very small for $k>k_{\text{max}}$ since $\boldsymbol{\Psi}_{in}$ will not have significantly high modes. The regularity of $\boldsymbol{\Psi}_{in}$ is also connected to the change of $c_m \ofxt$ over $\partial \Omega_i$. In most of the applications (e.g., forced convection) $c_m \ofxt$ varies regularly over the interfaces, and thus the correctors $\boldsymbol{\chi}_n$ vary smoothly (and slowly) over $\partial \Omega_i$. Therefore, we can also assume that there exist a $k_{\text{max}}$ such that:
\begin{equation}
    \label{eq::corr_kmax}
    \boldsymbol{\Psi}_{ikn} = 0
    \quad \mbox{for}\, {k>k_{\text{max}}}
\end{equation}
and, consequently,
\begin{equation}
    \label{eq::equilibrium_modes}
     c_{ik} = \favre{c_m}
     \quad 
     \mbox{for}\, {k>k_{\text{max}}}
\end{equation}
which means that for sufficiently large $k$ the modes are in equilibrium with the average field. Therefore, a more sensible approximation of the truncated terms would be a scaling of $\beta_i$ (and thus $\beta_m$) to account for the removed modes compared to simply rescaling $\beta_{ik}$. In practice, one would then have  new truncated immobile capacities $\beta_i^{\text{tc}}$ and mobile capacities $\beta_m^{\text{tc}}$  defined as:

\begin{equation}
    \label{eq::beta_tc}
    \beta_i^{\text{tc}} = \beta_i - \sum \limits_{k>k_{\text{max}}} \beta_{ik}, \quad \beta_m^{\text{tc}} = \beta_m + \sum \limits_{i=1}^{N_i}  \left( \beta_i - \beta_i^{\text{tc}} \right)
\end{equation}

Thus, the system behaves as if the mobile region was larger and the immobile regions were smaller (in terms of volumetric occupation, not geometrical parameters). This is simply due to the fact that we assumed that the dynamics of modes $k>k_{\text{max}}$ in the immobile regions is completely determined by the mobile region. 

\subsection{The multi-rate model of Haggerty \& Gorelick: the leading order approximation}

The original Multi-Rate Mass Transfer (MRMT) model proposed by Haggerty \& Gorelick \cite{Haggerty1995} can be obtained as
a special case of our general formulation.
More specifically, their model can be considered as a leading order approximation for $\favre{\psi_i}$, which results in the system: 
\begin{equation}
    \label{eq::mrmt_HG}
    \mbox{(MRMT)}\qquad
    \begin{dcases}
    &\beta_m \ddt{\favre{c_m}} +  \sum \limits_{i=1}^{N_i} \sum \limits_{k=1}^{\infty} \beta_{ik}\ddt{ c_{ik}}= - \div \mathbf{J}_{m,\text{eff}}   \\
    &    \ddt{c_{ik}} = \lambda_{ik}\left(c_{ik} - \favre{c_m} \right), \quad \begin{matrix} i=1,\dots,N_i,\\k=1,\dots,\infty\end{matrix}
    \end{dcases}
\end{equation}

Therefore, the model of Haggerty \& Gorelick is obtained under the approximation that the concentration at the interface between each immobile region and the mobile region is uniform and equal to $\favre{c_m}$. This is acceptable for systems where the mobile region is approximatively in local equilibrium at the microscale. This can be the case of a well-mixed concentration in the mobile region.
\subsection{Computation of $\beta_{ik}$ and $\lambda_{ik}$}
Coefficients $\beta_{ik}$ and $\lambda_{ik}$ do not depend in any way on the interface concentration $c_m\ofxt$ and, following our approach, they bear no dependence on the transport processes happening in the mobile region. Therefore, they can be calculated exactly using only geometrical shape and material properties of the immobile regions as input.

Table \ref{table::lambdabeta_simple} shows the expression of $\lambda_{ik}$ and $\beta_{ik}$ for a set of simple geometries. Clearly, our coefficients match those proposed by Haggerty $\&$ Gorelick \citep{Haggerty1995}, except for a factor $\beta_m$ in  $\beta_{ik}$, which is consistent with our formulation since we do not divide the equation for $\favre{c}_m$ by $\beta_m$.

\begin{table}[h!]
\centering
\begin{tabular}{c c c } 
 \hline
  Geometry & $\lambda_{ik}$ & $\beta_{ik}$  \\  
 \hline\\
 1d Layer & \(\displaystyle \left(2k -1 \right)^2\pi^2 \frac{\mathcal{D}_i}{4L_i^2}\) & \(\displaystyle\frac{8}{\left(2k -1 \right)^2 \pi^2}\) \\&&\\
 Cylinder & \(\displaystyle \zeta_k^2 \dfrac{\mathcal{D}_i}{R^2}\) & $\displaystyle \frac{4}{\zeta_k^2}$ \\
 &&\\
 Sphere & $\displaystyle k^2\pi^2 \frac{\mathcal{D}_i}{R^2}$ & $ \displaystyle \frac{6}{k^2\pi^2}$ \\[2ex]
 \hline
\end{tabular}
\caption{Evaluation of $\lambda_{ik}$ and $\beta_{ik}$ for simple geometries for which there is an analytical solution of the unsteady diffusion equation (see \citep{Crank1975} for details). Here $L$ represents half the length of the layer (the domain in the layer goes from $-L$ to $L$), $R$ is the radius of the sphere or cylinder and $\zeta_k$ is the $k$-th zero of the zero-order Bessel function of the first kind.}
\label{table::lambdabeta_simple}
\end{table}

In this approximation. coefficients $\lambda_{ik}$ and $\beta_{ik}$ have the same meaning as in Haggerty $\&$ Gorelick, where $\lambda_{ik}$ plays the role of exchange rate between $c_{ik}$ and $\psi_{ik}$. As demonstrated in table \ref{table::lambdabeta_simple}, $\lambda_{ik}$ is a function of geometrical dimensions and material properties through $\lambda_{ik}$ =$\lambda_{ik}^{\star}\mathcal{D}_i/L_i^2$, where the dimensionless eigenvalue depends on the shape of the immobile region only.  On the contrary, $\beta_{ik}$ is a dimensionless weight that depends only on the class of geometrical shapes. 

\section{Beyond  classic MRMT}

While \cref{eq::mrmt_general} allows to easily recover the standard MRMT model in the limit of equilibrium concentration in the mobile region, the presence of a mixed derivative makes its physical interpretation rather cumbersome. Furthermore, such term can introduce instabilities in numerical solution algorithms.

To this end, it is useful to rewrite \cref{eq::Mdot_general} using the integral form of the exchange rate:
\begin{equation}
    \label{eq::Mdot_int}
    \Mdot_i = \frac{1}{V}\int \limits_{\partial \Omega_i} \mathcal{D}_i \frac{\partial \theta_i}{\partial n}\text{d}S +   \sum \limits_{k=1}^{\infty} w_{ik}c_{ik}\frac{1}{V}\int \limits_{\partial \Omega_i} \mathcal{D}_i \frac{\partial \phi_{ik}}{\partial n}\text{d}S
\end{equation}

Integrating the eigenvalue \cref{eq::phi_eigen} over $\Omega_i$, we can obtain the following relation for the eigenvalues:
\begin{equation}
    \label{eq::lambda_ik}
    \lambda_{ik} = \frac{1}{m_{ij}}\int \limits_{\partial \Omega_i} \mathcal{D}_i \frac{\partial \phi_{ik}}{\partial n}\text{d}S
\end{equation}

Expanding the first term on the right hand side of \cref{eq::Mdot_int} leads to:

\begin{equation}
    \label{eq::Mdot_theta}
    \frac{1}{V}\int \limits_{\partial \Omega_i} \mathcal{D}_i \frac{\partial \theta_i}{\partial n}\text{d}S = \frac{1}{V}\int \limits_{\partial \Omega_i} \mathcal{D}_i \frac{\partial \psi_i}{\partial n}\text{d}S - \frac{1}{V} \sum \limits_{k=1}^{\infty} w_{ik}\psi_{ik}\int \limits_{\partial \Omega_i} \mathcal{D}_i \frac{\partial \phi_{ik}}{\partial n}\text{d}S  = 
    - \sum \limits_{k=1}^{\infty} \beta_i\beta_{ik}\lambda_{ik}\psi_{ik},
\end{equation}
where we employed \cref{eq::Psi_boundInt} on the right-hand-side.
Then, substituting expansion \cref{eq::psii_exp} results into:
\begin{equation}
    \label{eq::Mdot_theta_2}
    \begin{split}
    \frac{1}{V}\int \limits_{\partial \Omega_i} \mathcal{D}_i \frac{\partial \theta_i}{\partial n}\text{d}S &= 
    -\beta_i\sum \limits_{k=1}^{\infty} \beta_{ik}\lambda_{ik} \left(\favre{c_m} + \sum \limits_{n=1}^{\infty} \boldsymbol{\Psi}_{ink}:\grad^n\favre{c_m}  \right)
    \end{split}
\end{equation}
The additional terms are consistent with the evolution equation for $c_{ik}$, so that the mobile-to-immobile fluxes are identical to the corresponding immobile-to-mobile flux regardless the number of terms retained in the expansions.

\subsection{Generalised Multi-Rate Transfer equations}
The complete set of equations \ref{eq::mrmt_general} can be therefore rewritten without mixed time-space derivatives as:
\begin{equation}
    \label{eq::GMRT}
    \mbox{(GMRT)}\qquad 
    \begin{dcases}
    \beta_m \ddt{\favre{c_m}} +  \sum \limits_{i=1}^{N_i} \beta_i \sum \limits_{k=1}^{\infty} \beta_{ik}\lambda_{ik}\left(c_{ik}-\favre{c_m} - \sum \limits_{n=1}^{\infty}\boldsymbol{\Psi}_{ikn}:\grad^n\favre{c_m} \right)= - \div \mathbf{J}_{m,\text{eff}},&  \\
        \ddt{c_{ik}} = \lambda_{ik}\left(c_{ik} - c^{\text{eq}}_{ik}\right),& \quad \begin{matrix} i=1,\dots,N_i,\\k=1,\dots,\infty\\\end{matrix}
    \end{dcases}
\end{equation}
where we defined the equilibrium concentration for term $k$ of region $i$ as:
\begin{equation}
    \label{eq::c_eq}
     c^{\text{eq}}_{ik} = \favre{c}_m + \sum \limits_{n=1}^{\infty}\boldsymbol{\Psi}_{ikn}:\grad^n\favre{c_m}
\end{equation}
This system of equations does not pose any significant issue for corrections up to the second order, since the order of the differential operators remains unchanged and no mixed derivatives arise. Physically, these additional terms change the equilibrium concentration at which $\partial c_{ik}/\partial t = 0$. 

\subsection{First order correction and drift flux approximation}

Retaining first order corrections in \cref{eq::GMRT} is equivalent to adding a drift like term to the standard multi-rate equation for the mobile region. The governing equations are given by:
\begin{equation}
    \label{eq::GMRT_1}
       \mbox{(GMRT-1)}\qquad 
    \begin{dcases}
    \beta_m \ddt{\favre{c_m}} +  \sum \limits_{i=1}^{N_i} \beta_i\sum \limits_{k=1}^{\infty} \beta_{ik}\lambda_{ik}\left(c_{ik}-\favre{c_m}\right)= - \div \mathbf{J}_{m,\text{eff}} + \sum \limits_{i=1}^{N_i} \beta_i \sum \limits_{k=1}^{\infty} \beta_{ik}\lambda_{ik} \boldsymbol{\Psi}_{ik1} \cdot\grad\favre{c_m},&  \\
        \ddt{c_{ik}} = \lambda_{ik}\left(c_{ik} - \favre{c_m} - \boldsymbol{\Psi}_{ik1}\cdot\grad\favre{c_m}\right),& \quad \begin{matrix} i=1,\dots,N_i,\\k=1,\dots,\infty\\\end{matrix}
    \end{dcases}
\end{equation}

For the special case in which the material microstructure does not vary in space and the flow field is macroscopically homogeneous (i.e., $\bm{u}_{\text{eff}} = \text{const}$),  $\boldsymbol{\Psi}_{ik1}$ does not depends on the spatial coordinates and we can define a drift velocity:
\begin{equation}
    \label{eq::drift_vel}
    \bm{u}_{\text{drift}} = - \sum \limits_{i=1}^{N_i} \beta_i \sum \limits_{k=1}^{\infty} \beta_{ik}\lambda_{ik} \boldsymbol{\Psi}_{ik1},
\end{equation}
and thus a new effective velocity:
\begin{equation}
    \label{eq::star_eff_v}
    \bm{u}^{\star}_{\text{eff}} = \bm{u}_{\text{eff}} + \bm{u}_{\text{drift}}.
\end{equation}
Therefore, the equation for the mobile region simply reduces to a standard advection diffusion equation, with an additional multi-rate reactive term:
\begin{equation}
    \label{eq::mob_drift}
     \beta_m \ddt{\favre{c_m}} + \grad \cdot \left( \bm{u}^{\star}_{\text{eff}} \favre{c}_m - \mathcal{D}_{\text{eff}} \grad\favre{c_m} \right)=   \sum \limits_{i=1}^{N_i} \beta_i\sum \limits_{k=1}^{\infty} \beta_{ik}\lambda_{ik}\left(\favre{c_m}-c_{ik}\right)
\end{equation}

\subsection{Physical considerations on $\boldsymbol{\Psi}_{ik1}$ }

\cref{eq::GMRT} is describing a reactive system where the equilibrium concentrations of the immobile regions are not the same as the concentration in the mobile region. Thus, in our model, the equilibrium point is shifted by the correctors, based on the gradients of $c_m$. This is a direct consequence of non-equilibrium at the microscale (i.e., within $\Omega$) and can be attributed (at least asymptotically) to the the flow field and to the existence of boundary layers, which effectively results in a different equilibrium concentration for each $c_{ik}$. Our approach based on the synergy between homogenisation theory and spectral decomposition provides a formal way to account for such non-equilibrium.

\begin{figure}[h!]
    \centering
    \includegraphics[width=0.85\textwidth]{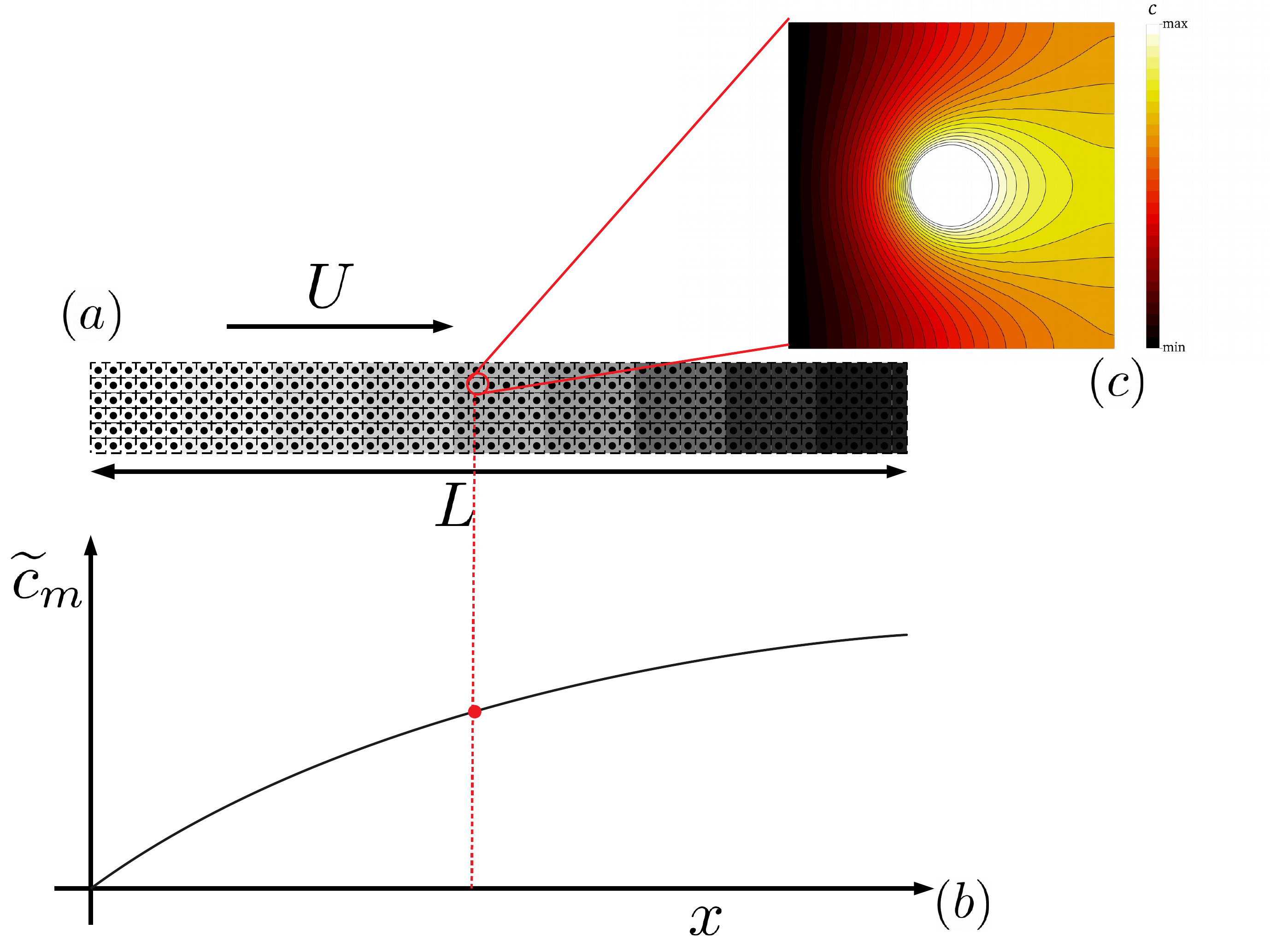}
    \caption{Illustration showing the state of a porous system composed of repeating cells with spherical inclusions at a time $t_0$. Here, a fluid moving with a uniform macroscopic velocity $U$ exchanges mass with a set of immobile regions at the same concentration (a), therefore $\favre{c}_m$ increases with $x$, with negative second derivative due to saturation (b). Panel (c) shows the expected contours for the local concentration around a spherical inclusion.  }
    \label{fig::MRMT_scalar}
\end{figure}

In order to understand the meaning of these corrector terms, it is useful to consider the toy case depicted in \cref{fig::MRMT_scalar}.

Such system is fundamentally monodimensional, and can be characterised by having:
\begin{equation}
    \label{eq::gradcm}
    \frac{\partial \favre{c}_m}{\partial x} \geq 0, \quad \forall x \in [0,L].
\end{equation} 

Now, we consider \cref{eq::cm_homo} at the first order:
\begin{equation}
    \label{eq::cm_first}
    c_m \ofxt = \favre{c}_m \ofxt + \boldsymbol{\chi}_1\ofx\cdot \grad \favre{c}_m
\end{equation}

Considering the local concentration in the mobile region, if all immobile regions have the same initial concentration, it follows that the local maxima will be locate at the interfaces as depicted in \cref{fig::MRMT_scalar}. Thus, considering that inequality \ref{eq::gradcm} holds, the same inequality holds for the $x$ component of the first order corrector $\chi_{1,x}$:

\begin{equation}
    \label{eq::chi1_constraint}
    \chi_{1,x}  > 0, 
\end{equation}
Therefore, if all components of $\grad \favre{c}_m$ are positive, all the components of $\boldsymbol{\chi}_1 $ are also positive. This positivity is transferred to $\boldsymbol{\Psi}_{i1}$ through \cref{eq::Psi_in} due to the properties of elliptic operators. While $\boldsymbol{\Psi}_{ik1}$ is not necessarily positive, the projection on the first eigenfunction  $\boldsymbol{\Psi}_{i11}$ is positive.

It is easy to demonstrate that this positivity property holds also when $\grad \favre{c}_m < 0$. 

Therefore, as illustrated in \cref{fig::MRMT_scalar}, a in a system with $\grad \favre{c}_m > 0$, the equilibrium concentration in the immobile regions will be larger than $\favre{c}_m$ due to the higher value of $c_m$ at the interface. On the contrary, in the case $\grad \favre{c}_m < 0$, this will be lower. 

While such argument was based on the analysis of a simple system, it is often valid for a large range of situations as, for example, in aquifer remediation and in many applications it is possible to guess the sign of the correctors by looking at the gradients. 

However, when different immobile regions have different initial conditions or the transfer in the immobile regions is strongly asymmetric, this positivity condition may be violated.

\subsection{Second order correction and diffusive flux approximation}

We now consider correction terms up to second order. Such term brings a second order differential operator into \cref{eq::GMRT}:
\begin{equation}
    \label{eq::GMRT_2}
       \mbox{(GMRT-2)}\qquad
    \begin{dcases}
    \beta_m \ddt{\favre{c_m}} +  \sum \limits_{i=1}^{N_i} \beta_i\sum \limits_{k=1}^{\infty} \beta_{ik}\lambda_{ik}\left(c_{ik}-\favre{c_m}\right)= - \div \mathbf{J}_{m,\text{eff}} - \sum \limits_{i=1}^{N_i} \beta_i \sum \limits_{k=1}^{\infty} \beta_{ik}\lambda_{ik} \left(\boldsymbol{\Psi}_{ik1} \cdot\grad\favre{c_m} +\boldsymbol{\Psi}_{ik2} : \grad\grad\favre{c_m} \right),&  \\
        \ddt{c_{ik}} = \lambda_{ik}\left(c_{ik} - \favre{c_m}\right) -\lambda_{ik} \left(\boldsymbol{\Psi}_{ik1} +\boldsymbol{\Psi}_{ik2}\cdot\grad\right)\cdot\grad\favre{c_m},& \begin{matrix} i=1,\dots,N_i,\\k=1,\dots,\infty\\\end{matrix}
    \end{dcases}
\end{equation}

Now, we can decompose tensor $\boldsymbol{\Psi}_{ik2}$ into hydrostatic and deviatoric components:
\begin{equation}
    \label{eq::theta2_deco}
    \boldsymbol{\Psi}_{ik2} = \text{dev}\left(\boldsymbol{\Psi}_{ik2}\right) + \frac{1}{3}\text{tr}\left(\boldsymbol{\Psi}_{ik2}\right)\mathbf{I}
\end{equation}

Where $\text{tr}\left(\boldsymbol{\Psi}_{ik2}\right)$ is the trace of $\boldsymbol{\Psi}_{ik2}$ and $\mathbf{I}$ is the identity tensor. We now introduce the diffusion coefficient arising from the conjugate transfer $\mathcal{D}_{\text{ct}}$:
\begin{equation}
    \label{eq::Dc}
    \mathcal{D}_{\text{ct}} = \sum \limits_{i=1}^{N_i}\beta_i\sum\limits_{k=1}^{\infty}\frac{\lambda_{ik}\beta_{ik}}{3}\text{tr}\left(\boldsymbol{\Psi}_{ik2}\right),
\end{equation}
which correspond to the second order correction arising in the case of macroscopically isotropic material with istropic immobile regions. 

Again, we make the approximation of homogeneous, isotropic material with macroscopically homogeneous velocity field so that $\mathcal{D}_{\text{ct}}$ does not depend on the spatial coordinate.
Under these approximations, the second order correction term becomes a purely diffusive contribution and we can thus define a new total diffusion coefficient:
\begin{equation}
    \label{eq::D_total}
    \mathcal{D}_{\text{tot}} =\mathcal{D}_{\text{eff}} - \mathcal{D}_{\text{ct}}
\end{equation}
Therefore, the equation for the mobile concentration simplifies to:
\begin{equation}
    \label{eq::mob_diff}
     \beta_m \ddt{\favre{c_m}} + \grad \cdot \left(\bm{u}_{\text{drift}} \favre{c}_m - \mathcal{D}_{\text{ct}} \grad\favre{c_m} \right)=   \sum \limits_{i=1}^{N_i} \beta_i\sum \limits_{k=1}^{\infty} \beta_{ik}\lambda_{ik}\left(\favre{c_m}-c_{ik}\right)
\end{equation}



\section{Summary of model parameters}
\label{S:summ}
Clearly, the the multi-rate series would be generally truncated at a desired accuracy.
All the correction terms arising in the formulation of the present model can be evaluated based on analytical or numerical analysis of the immobile and mobile regions. All such parameters can be evaluate \emph{a priori} and do not require additional computation when solving the macroscopic problem. 

Specifically, two parameters are independent on the flow and geometry in the mobile region:
\begin{itemize}
    \item[$\lambda_{ik}$] : these are simply the eigenvalues corresponding to the homogeneous eigenproblem in the immobile region.
    \item[$\beta_{ik}$] : these weights can be calculated similarly to $\lambda_{ik}$, from the solution of the eigenproblem. Once the eigenfunctions are known, $\beta_{ik}$ is given by: $\beta_{ik} = (\int_{\Omega_i} \phi_{ik}\text{d}V)^2/(V_i \int_{\Omega_i} \phi_{ik}^2\text{d}V)$
\end{itemize}
These are the same parameters of standard multi-rate models.
Furthermore, there ore other parameters that require the solution of a cell problem in the mobile region and therefore, that bring information regarding the interplay of conjugate transfer and transport in the mobile region. Such terms make use of the correctors $\boldsymbol{\chi}_{in}$ obtained from homogenisation theory.
\begin{itemize}
    \item[$\boldsymbol{\Psi}_{ikn}$]: Projection of the function $\boldsymbol{\Psi}_{in}$ on the eigenfunction $\phi_{ik}$ scaled with the norm of $\phi_{ik}$. Clearly, the number of these parameters equals the number of terms in the multi-rate expansion but one can exploit some knowledge of the microstructure to simplify their expression.
\end{itemize}

Other quantities we introduced, like $\mathcal{D}_{\text{ct}}$, $c^{\text{eq}}_{ik}$ or $\bm{u}_{\text{drift}}$, can be obtained from the other parameters.

It is worth to notice that, as it is often suggested for the standard MRMT, it is possible to consider each of the parameter as unknown and obtainable (for example) trough inverse analysis or data fitting. In this case, while the details of the derivation of $\lambda_{ik}$, $\beta_{ik}$ and $\boldsymbol{\Psi}_{ikn}$ become irrelevant, it is still crucial to remember that all the physics of non equilibrium is contained in $\boldsymbol{\Psi}_{ikn}$.

\section{Conclusions}
\label{S:end}
In this paper we propose a novel approach to derive the multi-rate mass transfer model that is different from that of the memory function or that of Haggerty $\&$ Gorelick. Our model is derived starting from the microscopic equations and it is parameter free, i.e., it is possible to directly evaluate all the closure parameters in a unique manner.
While our method agrees with previous results obtained by Haggerty $\&$ Gorelick,  it also contains their multi-rate model as a special case and allows extension to non-equilibrium situations, where the concentration in the mobile region is not uniform.
Especially, when homogenisation techniques are employed to evaluate the effective transport in the mobile region, our method provides an exact framework for the upscaling of the conjugate transfer problem, the accuracy of which is given by the terms retained from the infinite series.

Our model predicts that additional arise in the governing equations of the multi-rate mass transfer when accounting for the effect of transport processes in the mobile region on the inter-region exchange. These terms are brought into the framework by the corrector equation resulting from homogenisation, which at the second order have the form of a drift and a diffusive contribution.

Furthermore, under the assumptions of isotropy and homogeneity these terms can be absorbed into the effective diffusivity and effective velocity, thus leaving the form of the governing equations in the mobile region unchanged. However, the concentration in each immobile region will now depend on high order spatial derivatives of the concentration in the mobile region.  

Despite the self-consistency of this model (all the parameters can be evaluated from first principles without calibration) and its completeness with respect to the initial hypothesis (we never introduced additional hypothesis or simplifications in the development of our formulation) there are still some significant phenomena that should be accounted for when modelling real systems. Some examples are:
\begin{itemize}
    \item Exchange between immobile regions.
    \item Multiple mobile regions with different mobility (e.g., fractures).
    \item Chemical reactions at interfaces.
    \item Multiphase flow, heat and mass transfer.
    \item Internal flow currents in the immobile regions.
\end{itemize}
Future works could focus on one or more of these topics to improve the range of applicability of this proposed model.

\section{Acknowledgements}
This work has been funded by the European Union’s Horizon 2020
research and innovation programme, grant agreement number 764531, "SECURe – Subsurface Evaluation of Carbon capture and storage and Unconventional risks".




\bibliographystyle{model1-num-names}
\bibliography{references}







 \appendix

\section{Homogenisation: evaluation of the first order corrector}
\label{S:appendix}
Homogenisation is a perturbative method that allows to separate the original multiscale problem into a hierarchy of problems acting at different scales. 
In the following, we show how an equation for the immobile concentration similar to equation \ref{eq::mc_1} can be obtained using homogenisation and how to calculate the first order corrector $\boldsymbol{\chi}_1$. The purpose of this Appendix is simply to illustrate the method applied to the current study. For a detailed and rigorous description of the homogenisation procedure see for example \cite{Auriault1995}.

The process starts defining an expansion parameter:

\begin{equation}
    \label{eq::bookkeeping}
    \varepsilon = \frac{R}{L} \ll 1
\end{equation}

and a microscopic scale $\bm{y}$ such that:
\begin{equation}
    \label{eq::yscale}
    \bm{y} = \frac{\bm{x}}{\varepsilon}
\end{equation}

Where $R$ is a characteristic length of the immobile regions and $L$ is a characteristic length at the macroscale. The field $c_m \ofxyt$ is then expanded in asymptotic series of $\varepsilon$:

\begin{equation}
    \label{eq::asympt}
    c_m \ofxyt = \sum \limits_{n=0}^{\infty} \varepsilon^n c_{mn}\ofxyt
\end{equation}

Spatial differential operators are expanded to account for the microscopic scale:

\begin{equation}
    \label{eq::diff_y}
    \grad = \grad_{\bm{x}} + \varepsilon^{-1}\grad_{\bm{y}}
\end{equation}

Here $\bm{y}$ represents the variation across the REV $\Omega$, while $\bm{x}$ is a coordinate on the macroscopic volume $\Omega_{\text{macro}}$. This splitting is also known as two-scale asymptotics.

Then, \ref{eq::asympt} and \ref{eq::diff_y} are substituted into equation \ref{eq::mc_1} and, retaining terms up to $\mathcal{O}\left(\varepsilon^{-2} \right)$:

\begin{equation}\label{eq::expandedDispersionstep3}
    \begin{aligned}
    &\eps^{-2} \bigg\{\ny \cdot ( \bm{u} c_{m0} - \mathcal{D}_m\ny c_{m0})\bigg\}+& \notag \\
    &\eps^{-1}\bigg\{ \nx \cdot ( \bm{u} c_{m0} - \mathcal{D}_m\ny c_{m0} ) -\ny \cdot [ \mathcal{D}_m(\nx c_{m0}+ \ny c_{m1})-\bm{u} c_1  ] +& \notag \\
    &\eps^0 \bigg\{ \frac{\partial c_{m0}}{\partial t}  - \nx \cdot [ \mathcal{D}_m (\nx c_{m0}+\ny c_{m1})] - \ny \cdot \mathcal{D}_m[(\nx c_{m1}+\ny c_{m2})] + \notag\\
    &\hspace{30pt}+ \ \nx \cdot( \bm{u}  c_{m1})\bigg\} 
     = \mathcal{O}(\eps)&  \\
    \end{aligned}
\end{equation}

The boundary conditions are of the second type, in agreement with equation \ref{eq::bc}:

\begin{equation}
    \label{eq::bch}
    \mathcal{D}_m\grad c_m = \mathcal{D}_i \grad c_i = - \eps \bm{f}_i \ofxyt, \quad \bm{x} \in \partial \Omega_i 
\end{equation}

Where the order $\eps$ is taken due to the scaling of $\bm{f}_i$ with the specific surface and the negative sign takes into account for the vectors normal to the surface. This can be considered as an approximation of ''slow flux", we are fundamentally assuming that the diffusion is dominant at the macroscale with respect to the inter-region flux.  

The boundary conditions are also expanded (and multiplied by $\eps^{-1}$ to account for the surface-to-volume ratio ).

\begin{equation}
    \label{eq::bcexp}
    \mathcal{D}_m \left[ \eps^{-2}\left(\ny c_{m0} \right) + \eps^{-1} \left( \ny c_{m1} + \nx c_{m0} \right) + \eps^0\left( \nx c_{m1} + \ny c_{m2} \right) \right] = - \bm{f}_i
\end{equation}

Matching the orders, we obtain the following:

\begin{itemize}
    \item [$ \mathcal{O}(\eps^{-2})$] 
    
    The first equations simply gives the independence of the leading order from $\bm{y}$ (required by homogenisation):
        \begin{equation}
            c_{m0} = c_{m0}\ofxt 
        \end{equation}
        
    \item[$\mathcal{O}(\eps^{-1})$] 
    
    This equation can be solved posing:
    
    \begin{equation}
        c_{m1} = \boldsymbol{\chi}_1 (\bm{y}) : \nx c_{m0}\ofxt
    \end{equation}
    
    Introducing the first order corrector $\boldsymbol{\chi}_1$, which then satisfies:
    
    \begin{equation}
        \begin{dcases}
        \ny \cdot \left[ \mathcal{D}_m \left( \mathbf{I} + \ny \boldsymbol{\chi}_i \right) - \bm{u}\boldsymbol{\chi}_1 \right] = \bm{u}, & \bm{y} \in \Omega_m\\
        \textbf{I} + \ny \boldsymbol{\chi}_1 = 0, & \bm{y} \in \partial \Omega_i
        \end{dcases}
    \end{equation}

  \item[$\mathcal{O}(1)$] Matching the orders and applying Favre averaging over $\Omega_m$, this results in an upscaled equation for $c_{m0}$:
  
  \begin{equation}
      \beta_m \ddt{c_{m0}} + \sum \limits_{i=1}^{N_i} \Mdot_i = - \nx \cdot \left[ \bm{U}c_{m0} - \mathcal{D}_m \left( \mathbf{I} + \int \limits_{\Omega_m}\frac{\ny \boldsymbol{\chi}_i}{V_m} \text{d}V \right) \nx c_{m0} \right]  
  \end{equation}
  
  Where $\Mdot_i = \int_{\partial \Omega_i} \bm{f}_i \cdot \bm{n}_i \text{d}S$ is the same as equation \ref{eq::flux_i}. Furthermore, $\bm{U} = \int_{\Omega_m} (\bm{u}/V_m) \text{d}V $.
\end{itemize}

This also allows us to obtain an expression for the effective diffusivity, and to connect volume averaging and homogenisation, as  $\favre{c_{m}} = c_{m0}$. 
This formulation can be seen as an alternative to spatial filtering for the mobile region, although it is possible to obtain the same correctors equations using the volume averaging method (see for example \cite{Whitaker1999a}).

\end{document}